# Scalable nanoimprint manufacturing of multi-layer hybrid metasurface device

*Shinhyuk Choi[1], Jiawei Zuo[1], Nabasindhu Das[1], Yu Yao[1]\* and Chao Wang[1,2,\*]*

*[1]School of Electrical, Computer and Energy Engineering, Arizona State University, Tempe, AZ 85287, USA*

*[2]Biodesign Center for Molecular Design and Biomimetics, Arizona State University, AZ 85287, USA*

---

\* Corresponding author.

  E-mail address: yuyao@asu.edu; wangch@asu.edu

**Abstract**

Optical metasurfaces, consisting of subwavelength-scale meta-atom arrays, hold great promise to overcome fundamental limitations of conventional optics. Scalable nanomanufacturing of metasurfaces with high uniformity and reproducibility is key to transferring technology from laboratory demonstrations to commercialization. Recently, nanoimprint lithography (NIL) has attracted increasing interests for metasurface fabrication because of its superior nanometer resolution, rapid prototyping and large-area manufacturing capability. Despite NIL demonstrations of single-layer metasurface, scalable fabrication of double- and multi-layer metasurfaces remains challenging. Here we leverage the nanometer-scale resolution and 3D pattern transfer capability of NIL to fabricate multi-layered metasurfaces for on-chip polarimetric imaging devices. Our process achieved sub-100 nm nanostructures, high alignment accuracy (translational error <200 nm; rotational error <0.02°), and good uniformity (<4 nm linewidth deviation) over >20 mm$^2$. This NIL-based, low-cost and high-throughput nanomanufacturing approach paves the way toward scalable production of a plethora of metasurface structures for ultra-compact optic and optoelectronic devices and systems.

## Introduction

Metasurfaces are capable of manipulating fundamental electromagnetic responses, i.e., phase, amplitude, spectral response, and polarization [1-3], at the subwavelength scale. They have shown great potential in addressing fundamental limitations of conventional bulky optical systems and realizing ultracompact optical devices and systems for many applications, such as holography, imaging, spectroscopy, beam shaping and steering, etc.[4-7] Despite significant progress in metasurface design and proof-of-concept laboratory demonstrations, scalable and cost-effective nanomanufacturing with good uniformity and reproducibility remains one of the major challenges that slows down commercialization of metasurface devices. Conventional prototyping nanofabrication methods, such as electron-beam lithography (EBL) or focused-ion beam (FIB), reply on pixel-by-pixel writing for precise nanopatterning but are not suitable for scalable manufacturing due to long writing time, high cost and reproducibility problems over large scale[8]. High-throughput semiconductor optical lithography technologies (such as deep-UV or extreme-UV lithography) are too expensive, and complex to operate for prototyping demonstrations. In comparison, nanoimprint lithography (NIL) is suitable for both prototyping demonstration and large-scale production of nanostructures as small as to sub-ten nanometer scale given its unique optical diffraction-free, parallel patterning capabilities [9-11]. Previously, NIL has been employed successfully in a wide range of optical applications, such as polarizers[12, 13], anti-reflection coatings [14], solar cells[15], etc. Scalable NIL fabrications of a single-layer metasurface structure have been demonstrated [16-19]; however, so far the demonstrated processes were only for geometrically simple, stand-alone, single-layer metasurface structures. Yet the realization of multi-layer dielectric, metallic or hybrid metasurfaces requires not only high-throughput nanopatterning, but also precise alignment and vertical stacking quality control.

Uniquely, NIL is capable of not only two-dimensional lithographical patterning but also three-dimensional transfer-printing. Here, we strategically establish a scalable method towards manufacturing of metasurfaces over multiple layers, by using NIL as a lithography and printing method, respectively, to create microscale subwavelength-thick microscale polarization filter arrays [20-22] for integration onto complementary metal-oxide semiconductor (CMOS) imaging sensors. As a proof-of-concept demonstration, multi-layer dielectric and metallic hybrid metasurfaces with dense features (sub-100 nm features with periods ~200 nm) were integrated at a high alignment accuracy (interlayer alignment error ~200 nm, rotation error < 0.017 degrees) and accurate dimensional control (linewidth standard deviation less than 4 nm). Uniquely, by codesign of the NIL process and the metallic metasurface structures, we demonstrate that a single-step NIL pressing can effectively create a 3D metasurface scaffold, therefore eliminating a number of processing steps and greatly reducing manufacturing cost and time. Moreover, the same NIL process produces a much smoother spacer layer through its in-situ planarization capability, thereby greatly suppressing optical scattering loss and accordingly improving the polarization extinction ratio (ER) of the polarization filters. In addition, multi-layered circular polarization (CP) filters targeting at different visible wavelengths, together with broadband linear polarization (LP) filters are spatially arranged into arrayed superpixels and implemented across the whole chip, enabling broadband polarimetric imaging and full-Stokes parameter analysis across visible wavelengths at a high accuracy. This successful multilayer NIL-metasurface codesign approach can be adapted to the fabrication of many other metasurface structures, enabling high-throughput scalable manufacturing of various metasurface devices for both efficient prototyping and large-scale production of ultra-compact chip-integrated optic and optoelectronic devices and systems.

**Results and Discussion**

**Scalable Manufacturing Design and Process**

Here we present a synergistic approach to co-design the multi-layered optical metasurfaces and their scalable NIL manufacturing process. Our exemplary polarimetric imaging system was a multilayered metasurface polarization filter array (MPFA) integrated onto a CMOS imaging sensor (Fig. 1a). The MPFA consisted of over 43,000 superpixels, each having four LP filter pixels and four CP filter pixels (Fig. 1b) to ensure accurate full-Stokes polarization measurement. The LP filters were based on vertically coupled double-layer gratings (VCDGs) with high LP extinction ratio (LPER) over a broad wavelength range (Fig. 1c)[22]. The CP filters were based on multi-layered chiral metasurface structures[20, 22], consisting of a Si metasurface acting as quarter wave plate (QWP) (Supplementary Fig. S1), a dielectric spacing layer, and VCDGs as LP filters (Fig. 1d). Overall, the MPFA was formed by two vertically aligned, functional layers, i.e., the Si metasurface layer and the VCDG layer (Fig. 1a-b). In the Si metasurface layer, each superpixel had 4 blank pixels (no nanopatterns, pixels 1 to 4, Fig. 1b) and 4 pixels made of Si nanostructures (pixels 5 to 8). In the VCDG layer, the grating polarizers were present in all 8 pixels, oriented along 0°, 45°, 90° or 135° in the 4 LP filters and all along 90° in the 4 CP filters (Fig. 1b). To achieve a broadband coverage in visible (450 to 670 nm), two sets of CP filters (VCDGs + Si metasurface) were designed, one for green-wavelength operation (510 to 600 nm, pixels 7 and 8, Fig. 1b), and the other (pixels 5 and 6) for both blue (450 to 510 nm) and red wavelengths (600 to 670 nm) (detailed designs parameters and simulation results in Supplementary Table S1 and Fig. S2). This design enabled a single-shot, full-Stokes polarimetric analysis and imaging over a broad bandwidth in visible wavelengths. Each of the LP and CP filter pixels is shared by neighbor superpixels, thus maximizing the amount of effective superpixels in case of manufacturing defects for optimal imaging resolution. As a proof-of-concept demonstration, we fabricated the MPFA on a transparent silica substrate and then integrated it onto a commercial CMOS imaging sensor via polymer-assisted wafer bonding. The process can be readily modified to directly integrate the metasurface onto CMOS chips for wafer-scale production.

Previously, we have developed an EBL-based process to fabricate the MPFA[22], and demonstrated dual-color full-Stokes parameter detection with a high accuracy and a large field of view [22]. However, the fabrication process required extensive EBL writing time, repeated film deposition, lift-off and etching. Furthermore, the silicon oxide ($SiO_2$) spacer on top of the silicon (Si) metasurface displayed a rough surface, which resulted in uneven Al grating surfaces in the VCDGs and limited device LPER and CPER[22].  Fundamentally different from EBL (Fig. 1f top, figures following orange arrows), here NIL (Fig. 1f bottom, following green arrows) was utilized first as a high-throughput, high-resolution lithography technology to produce Si metasurface gratings, and then used as a three-dimensional surface topography replication process to print the VCDG grating scaffold in resists, which replaced the spacer layer in EBL fabrication. Thermal NIL was chosen for the Si metasurface fabrication for its simplicity, and UV-NIL was carried out for VCDGs by optically aligning a transparent mold to the Si metasurface and pressing the mold into a resist with an optical index comparable to $SiO_2$. Here Moiré patterns were created on both of the two NIL molds to achieve a high overlay accuracy over the patterned area. Moreover, the UV-NIL not only effectively produced the 3D VCDG scaffold, but also eliminated multiple manufacturing steps and planarized the resist despite underlying protruding topography from Si metasurface. Noticeably, NIL is many times faster than EBL when a mold is available (Supplementary Table S2), and its high throughput advantage can be further manifested when scaling to even bigger areas, reducing cost, and improving throughput in manufacturing (Supplementary Table S3). Therefore, such a new, simplified process simultaneously reduced processing complexity, improved the MPFA performance, and enabled scalable device production.

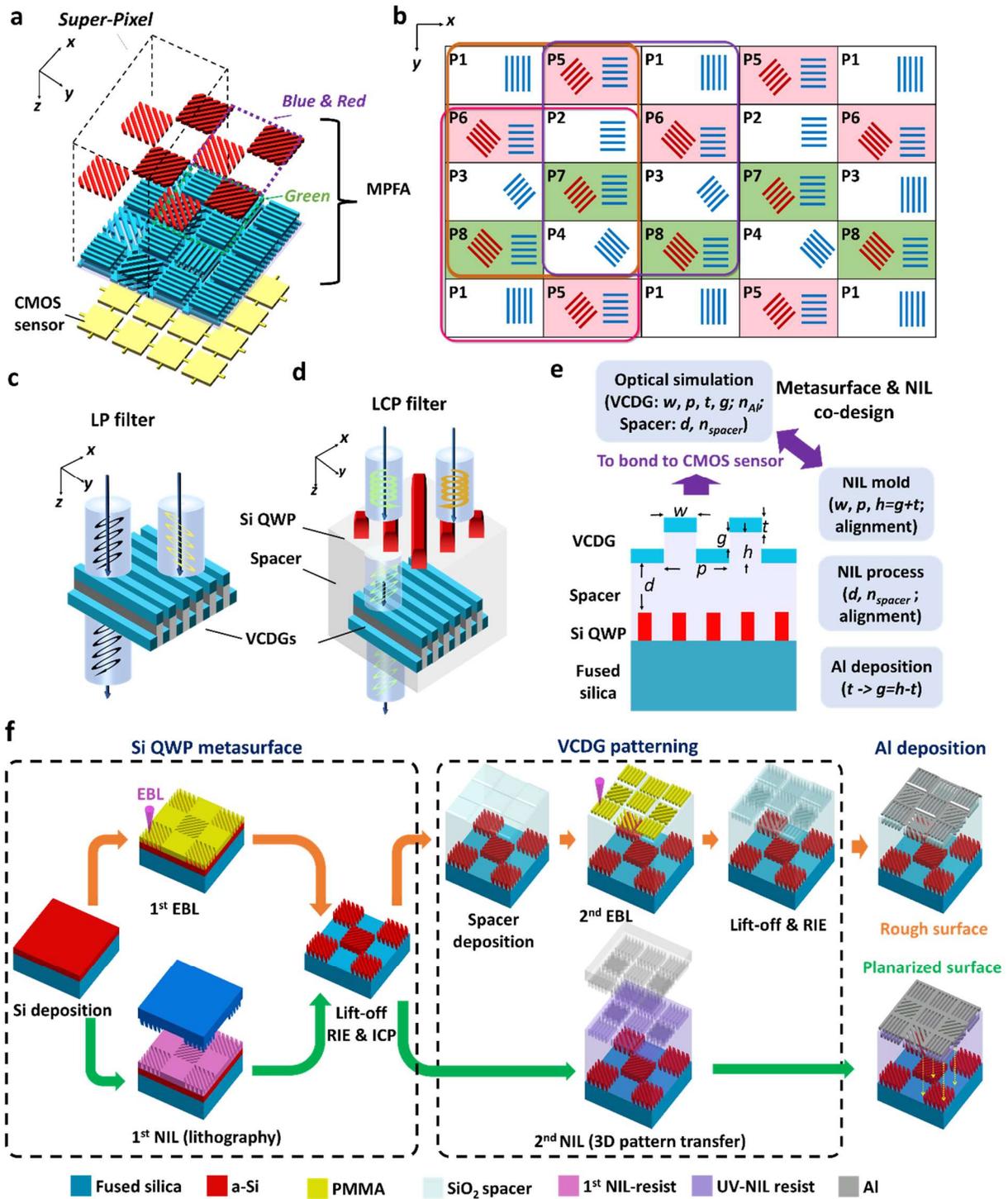

**Fig.1: Conceptual designs of scalable NIL manufacturing for multi-layer metasurface polarization filter array (MPFA). a,** Illustration of integrating CMOS imager with broadband MPFAs consisting of a layer of Si metasurface as quarter waveplate (QWP) and a layer of VCDGs as polarizers. Here two CP designs targeting green and blue/red spectra (indicated by arrows) are

incorporated for the Si metasurface structures. **b**, Illustration of the arrangement of each metasurface polarization filters (pixels) within super-pixels (rectangular boxes). Here P1-P4 have only VCDGs transmitting 0°, 90°, 45°, 135° LP light, respectively. P5-P8 are chiral metasurfaces constructed by Si QWP and VCDGs in each pixel that transmit RCP, LCP in red and blue color range (P5 and P6, red shaded) and green color range (P7 and P8). **c**, Schematic illustration of VCDG to pass light polarized along x axis but block light polarized along y axis. **d**, Schematic of multi-layered CP filter transmitting LCP but blocking RCP incoming light. **d**, A co-design concept to produce the VCDGs on Si metasurface structures based on NIL. Here the structural geometries and processing conditions are designed for optical performance. **e**, Schematics showing the EBL (top, following orange arrows) and NIL (bottom, following green arrows) based fabrication processes for MPFAs. Here a 1st NIL replaces EBL for the fabrication of Si metasurface, and a 2nd UV-NIL creates a nanostructured scaffold to be converted into VCDGs after Al evaporation.

To further improve the manufacturing throughput and device performance, we identified key design parameters closely relevant to NIL processes and crucial to performance of Si QWP and the VCDGs (Fig. 1e, Supplementary Table S1). At the Si QWP metasurface level, dense gratings of 180 nm or 297 nm in period and linewidths of 80 or 100 nm are needed for CP polarization filters of green or blue/red operational wavelengths (Supplementary Fig. S3). The simulation results indicate that small variations of the Si grating linewidth only slightly modulate the optimal CPER values and the peak wavelengths, showing a high design tolerance. At the VCDG level, the grating period ($p$), controlled by the NIL mold structure design, and width ($w$), determined by UV-NIL and subsequent processing conditions, both have strong influence on the LPER and optical transmission. The duty cycle ($w/p$) was designed at 50% with a tolerance of ±20% to control the LPER to over 1000 in visible (Supplementary Fig. S4) [22]. The vertical gaps between the Aluminum (Al) double gratings ($g$) was determined by the designed mold height and experimentally optimized Al thicknesses to maximize the LPER. Further, the spacing between the VCDGs and the Si nanostructures ($d$) also strongly affects wavelength ranges to achieve the best double-layer

MPFA performance (Supplementary Fig. S5), as it affects the phase accumulation of electromagnetic waves travelling between the two layers upon reflection.

**Silicon metasurface fabrication**

The silicon metasurface-based, microscale QWP array was fabricated using thermal-NIL with a NIL mold fabricated on a thermal $SiO_2$-coated silicon wafer (Fig. 2a). We first made the NIL mold by EBL patterning, chromium (Cr) hard mask deposition and liftoff, reactive ion etching (RIE) of $SiO_2$, and Cr stripping (Methods section and supplementary Fig. S6). Then we fabricated the Si metasurface from α-Si thin film deposited by chemical-vapor deposited (CVD) using a tri-layer pattern-transfer scheme. Here thermal NIL was performed on a film stack, including a bottom polymethyl methacrylate (PMMA) layer, a mid-layer evaporated $SiO_2$ film, and a top-layer thermoplastic NIL resist. Then a series of RIE with oxygen plasma or $CHF_3$ plasma was employed to selectively etch polymer (PMMA and NIL resist) or $SiO_2$, respectively, to transfer the NIL-patterned grating features to the film stack. Lastly we performed Cr deposition, Cr mask liftoff, and RIE of $SiO_2$/Si films to complete the fabrication of Si metasurface layer. Compared to a single-layer resist, the tri-layer film stack could effectively produce high-aspect-ratio nanostructures with improved patterning uniformity over a large area by reducing dimensional distortion and facilitating minimally defective Cr liftoff [23, 24], thus favorable for subsequent high-fidelity pattern transfer to the underneath Si metasurface gratings. Additionally, the grating linewidth could be adjusted within a range of about 30 nm by modulating the etching time of the top thermoplastic NIL resist, offering additional flexibility in controlling structure dimensions. We measured the linewidths of Si metasurface gratings from scanning electron microscopy (SEM) images at different locations of the sample (Fig. 2b), and found the standard deviations (SDs) less than 4 nm for all pixel designs (Supplementary Table S4), confirming good uniformity of the NIL process for large-scale nanomanufacturing of Si metasurface.

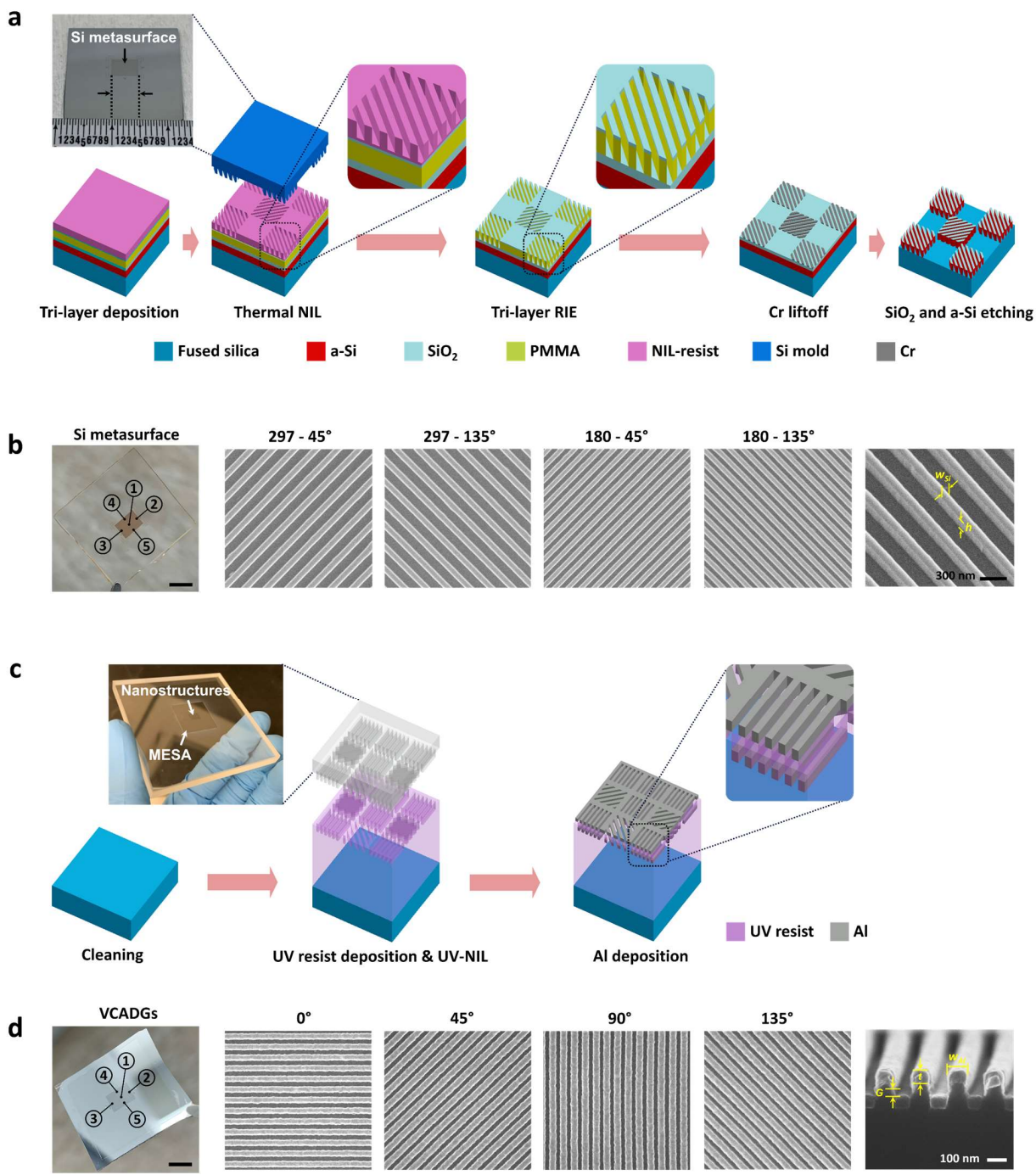

**Fig. 2: Schematics and fabrication results of Si and VCDG metasurface structures. a**, Fabrication process of Si metasurface via thermal NIL on a tri-layered resist structure, including PMMA (bottom), SiO$_2$ (middle, evaporated) and thermoplastic NIL resist (top). The key

processing steps include NIL, RIE of the tri-layer stack, Cr evaporation and liftoff, $SiO_2$ RIE, Cr stripping, and Si RIE. The top-left insert optical image shows the Si metasurface NIL mold made in a Si wafer. **b**, Fabricated Si metasurface structures on fused silica. Left: optical image of a fabricated Si metasurface chip (scale bar: 5mm). Middle: top-view SEM images of unit pixel arrays. Right: side-view SEM image of Si gratings (scale bar: 300 nm). **c**, Illustration of fabrication process of VCDGs by UV-NIL followed by Al evaporation. The top-left insert optical image shows the VCDG mold made in fused silica. **d**, Fabricated VCDG grating structures. Left: optical image of VCDG gratings on a silicon sample for structural inspection (scale bar: 5 mm). Middle: top-view SEM images of unit pixel arrays. Right: cross-sectional view of fabricated VCDGs after Al deposition (scale bar: 300 nm) with the key geometrical dimensions highlighted (*g*: gaps between the two sets of gratings; *t*: Al thickness; *w*: grating width). In Figures b and d, five areas are randomly chosen from the chips to examine the fabrication uniformity: ① Center, ② Right, ③ Left, ④ Top, and ⑤ Bottom. The measured structural dimensions are given in Supplementary Table S4.

**NIL 3D Scaffolding for metallic grating metasurface (VCDG)**

The VCDG microscale linear polarizer array was fabricated using UV-NIL with a NIL mold fabricated on a transparent fused silica wafer (Fig. 2c) for precise vertical stacking and alignment of the VCDG layer onto Si metasurface layer. The VCDG mold fabrication process was similar to that for the Si metasurface NIL mold, with more details in Methods section. Briefly, we intentionally created the nanostructured mold patterns on an etched mesa (Supplementary Fig. S7) to further improve the NIL pressure, which was beneficial to improve printing quality. To examine the VCDG NIL process, we coated a Si wafer with an acrylate-based UV-curable resist and performed UV-NIL on the sample. The UV-NIL transferred the nanograting patterns into the UV resist (optically $SiO_2$-like, Supplementary Fig. S8) that was more resistant to oxygen plasma damage or heating in metal deposition than organic polymers. Therefore, the UV NIL resist served perfectly as a rigid, transparent VCDG scaffold, allowing the completion of the VCDGs fabrication by a subsequent simple Al deposition (thickness $t$) to produce desired vertical gaps ($g=h-t$) (Fig. 1e). Essentially, this single NIL-based pattern transfer printing step replaced multiple manufacturing steps otherwise needed for EBL fabrication (Fig. 1f), including $SiO_2$ spacer deposition, EBL writing, plasma descum, Cr deposition, liftoff, $SiO_2$ dry etching, and Cr removal, therefore greatly improving the throughput. We took SEM images at five randomly selected locations across the chips (Fig. 2d), measured the VCDG linewidths, and determined the linewidth SDs were less than 2 nm (Supplementary Table S4). The linewidth of angled VCDGs (116 nm for 45° and 135°, respectively) was found slightly different from vertical and horizontal gratings (109 nm for 90° and 0°, respectively), mainly attributed to linewidth difference in the VCDG mold during EBL-based pattern generation and writing process[25], but these were still within the acceptable range for VCDG grating polarizers (Supplementary Fig. S4). Further, we optimized the Al film deposition conditions to minimize the roughness (Supplementary Fig. S9), experimentally analyzed the impact of Al thickness (t) on the optical performance of VCDGs (Supplementary Fig. S10) for optimal LPER, and chose here to have $t$=80 nm.

**Vertical alignment and integration**

Multi-layer metasurface fabrication usually faces two major challenges: 1) alignment between the different layers, and 2) the impact of existing surface topography on the subsequent fabrication. Here in our exemplary MPFA demonstration (Supplementary Fig. S11), the alignment strategy was carefully designed to enable sub-micrometer overlay accuracy for vertical stacking of the VCDG and Si metasurfaces. In addition, UV-NIL was used to produce a 3D VCDG scaffold that not only produced uniform nanostructures but also in-situ planarized the surface topography from Si metasurface.

To achieve submicron optical alignment, we designed and fabricated interference-based

Moiré patterns [27-30] (Fig. 3a) on both NIL molds for Si metasurfaces and VCGDs, respectively. Here, two sets of gratings with slightly different periods acted as Moiré marks to produce interference patterns with a period $P_{fringe}$, calculated as $P_{fringe} = P_1 \cdot P_2/(P_2 - P_1)$. Therefore, the misalignment between the two metasurfaces ($\Delta = G_i - G_j$) could be made much smaller than the visualized Moiré fringe offset ($s$) as $\Delta = s(P_2 - P_1)/P_1 \ll s$ when $P_2 \sim P_1$, thus resulting in nanometer-scaled alignment accuracy. When in good alignment, Moiré fringe minima were clearly positioned next to our designed small squares and crosses on two metasurface layers that served as alignment indicators in each of the alignment mark groups (e.g., $AM_1$ and $AM_2$, Fig. 3a). Because the thick NIL resist spacer layer blocks electron beam signals but allows optical visualization, we chose optical microscopy to measure the alignment errors ($\Delta_1$, $\Delta_2$, $\Delta_3$, and $\Delta_4$). The average alignment errors were found below 200 nm in both x- and y-directions within the mm-scaled structure (Supplementary Table S5), satisfying required accuracy in our MPFA design (~1.6 µm). The Moiré marks can be designed to achieve much higher overlay accuracy by engineering the optical scanning, stage, and control systems, e.g., sub-10 nm overlay is routinely achieved on ASML scanners using fundamentally similar interferometric marks for larger-scale production. Nevertheless, the Moiré alignment method allows future integration of metasurface structures with reduced pixel sizes, e.g. to submicron with our demonstrated NIL capability or even smaller on more advanced systems, thus further improving the imaging sensor pixel density.

On the other hand, the surface topography resulting from the selectively fabricated Si metasurface strongly affects subsequent VCDG fabrication. As observed in our previous demonstration based on EBL processes[22], vacuum deposition of $SiO_2$ films as a spacer layer could not completely flatten the substrate surface. As a result, VCDGs fabricated on top of this rough $SiO_2$ spacer layer would suffer from optical losses and accordingly low LPER. As a contact-based nanopatterning technology, NIL process typically prefers a flat substrate for high-fidelity, low-defect pattern transfer[26], because existing nanostructures could disrupt the resist flow, therefore trapping air bubbles and creating defects. To minimize the surface roughness and maximize the fabrication yield, here uniquely UV-NIL is utilized to produce a 3D VCDG scaffold, which readily functions as a template to complete VCDG fabrication through an Al evaporation and simultaneously acts as a dielectric spacer layer between the Si and VCDG metasurfaces. Therefore, this method not only eliminated complex fabrication steps, but also simultaneously planarized the substrate surface to have a much reduced root mean square roughness (~1.2 nm, extracted from a 1 $\mu m^2$ flattened area without gratings, Fig. 3c). Clearly the in-situ planarization, which was attributed to the effective resist filling owing to the low viscosity of UV resist layer, allowed us to faithfully produce nanograting scaffold from the NIL mold. In comparison, the $SiO_2$ spacer vacuum-deposited on Si metasurface for EBL-fabricated chips displayed a wobbling surface (roughness ~15.6 nm, Fig. 3c). Indeed, the Al VCDGs formed on the NIL scaffold (Fig. 3d) were found much flatter and smoother than those fabricated by EBL. Further, we took SEM images at five randomly selected locations to measure the linewidths of VCDGs overlaid on Si metasurface, and determined the SDs to be less than 4 nm (Supplementary Table S6), indicating a high structure uniformity comparable to that of single-layer Si metasurface (Supplementary Table S4).

We also performed optical spectroscopic measurements of the integrated multi-layer metasurfaces and compared the performance of EBL and NIL-fabricated MPFA samples (Fig. 3e, and Fig. 3f) [15]. Notably, the maximum CPER of the NIL-fabricated MPFA was ~10 times and ~4 times better at blue and red color wavelength ranges (~20 and 80, respectively) than that of EBL-fabricated device (~2 and ~20 [22]). The improved CPER is attributed to greatly enhanced LPER of the VCDGs over a broad visible wavelength range fabricated by NIL than that by EBL (Supplementary Fig. S12). In addition, the peak wavelength where the max CPER is achieved was blue-shifted for the NIL-fabricated device, due to its thicker spacer layer (520 nm) than that of EBL device (400 nm), which is consistent with simulation results (Supplementary Fig. S5). The transmission efficiencies were much lower than the designed value, attributed to the fact that VCDG mold had rounded trench edges after timed-RIE in amorphous silica, which subsequently produced rounded shoulders in the VCDG scaffold and resulted in overhangs on the top Al gratings. Such structural modulation could therefore lower the transmission intensity (Supplementary Fig. S10). The optical transmission proved sufficient in this work for polarimetric imaging and accurate Stokes parameter analysis, but can be improved if necessary in the future by optimizing the NIL mold fabrication process to improve the straightness of VCDG gratings and minimize the structure rounding at the grating foot. For example, this can be accomplished through optimization of etching recipe and film stack to pattern the mold nanostructures in a film (e.g. $SiO_2$) selectively over an underlying etch-stop layer (Supplementary Fig. S13).

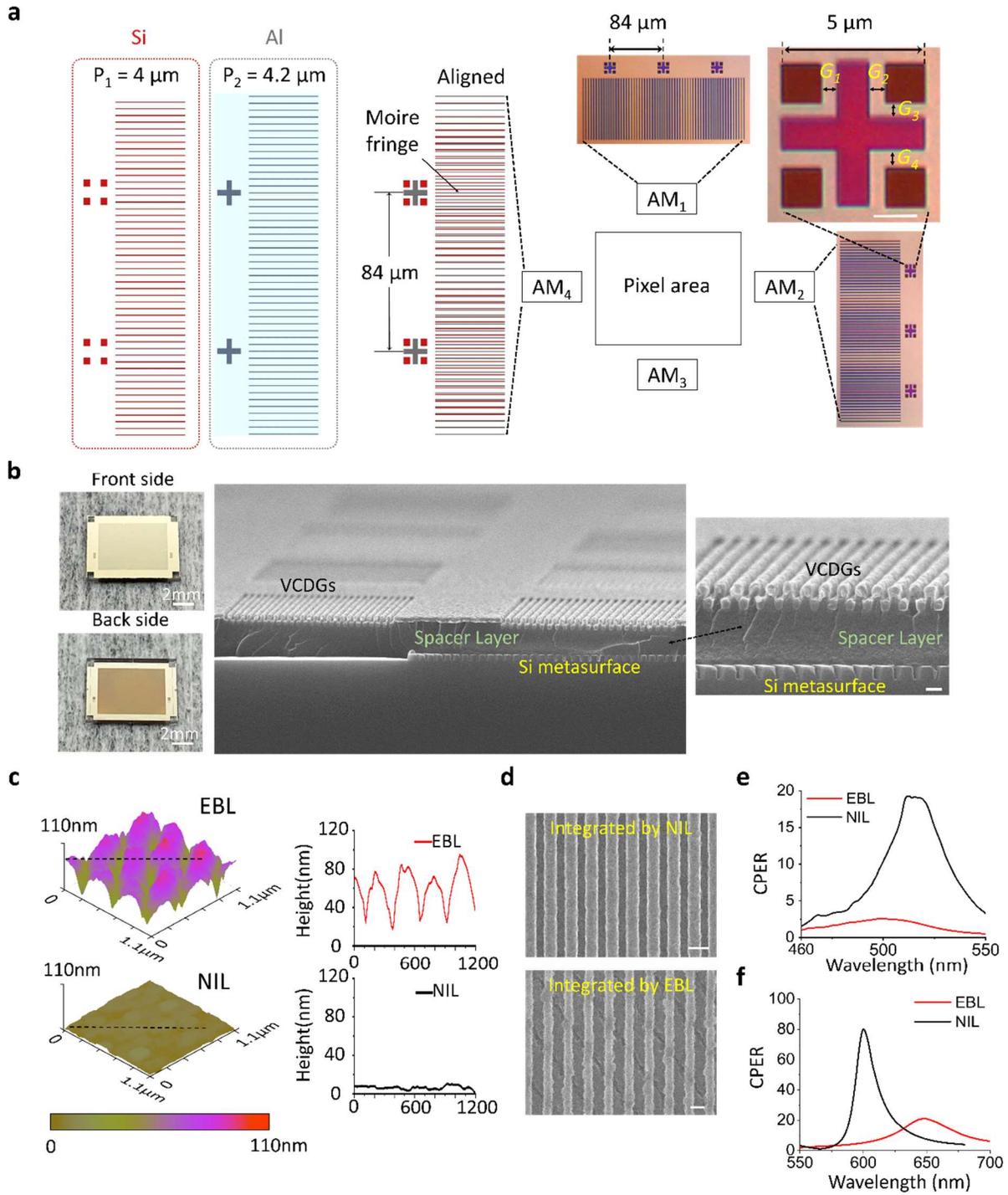

**Fig. 3: Integration and characterization of multi-layer metasurface chips for polarimetric imaging. a**, Moiré fringe-based alignment technique to achieve nano-scale overlay accuracy. Optical images of the top and right alignment marks (AM1 and AM2) showed clear interference patterns aligned to designed cross-square marks. The images of cross-square marks were used for alignment accuracy analysis (scale bar: 5μm). **b**, Integrated multi-layer metasurface chip. Left:

Optical images of the diced chip (top: VCDGs-side up; bottom: α-Si side up. Scale bar: 1mm) Middle and right: representative cross-sectional SEM image of integrated multi-layer metasurface structures (scale: 1 µm for middle image, and 200 nm for right image). **c**, Comparison of AFM images (left) and surface roughness profile (right) of the spacer layer of double layer metasurfaces fabricated by EBL (top) and NIL (bottom). **d**, Comparison of SEM images of EBL (top) and NIL (bottom) fabricated double layer metasurfaces. **e, f**, Comparison of characterized CPER of LCP filters fabricated by NIL (black curve) and EBL (red curve): **e**, at 460-550 nm, and **f**, at 550-680 nm, respectively.

## Imaging sensor integration and characterization

The integrated multi-layer MPFAs were diced (7.2 mm × 5.6 mm), optically aligned to the edges of a commercial CMOS sensor (IMX477) on mask aligner, and bonded with UV-curable polymer, as shown in Fig. 4a and 4b (details in Methods section). This alignment translational error was on the micrometer scale and the rotational error was about 0.02°, constrained by the lack of more accurate alignment marks (e.g. Moiré patterns) on the CMOS imaging sensors. To further minimize this alignment errors, one can design the layouts of CMOS imaging sensor and the metasurfaces with interferometric Moiré patterns, similar to what we demonstrated for high-accuracy alignment of multi-layer metasurface structures in the previous section. We characterized the bonded metasurface polarimetric imaging sensor (or Meta-PolarIm) to determine its instrument matrix $A$ at different wavelength bands, i.e., blue (480-520 nm), green (530-570 nm), and red (580-620 nm), respectively (see the method section for detailed procedures to obtain the instrument matrix A). Thus, the Stokes parameters of any unknown input polarization state $S$ can be obtained using $S = A^{-1}I$, where $I$ represents the intensity vector obtained by all 8 pixels in each super pixel of Meta-PolarIm[22]. We measured the eight polarization states (supplementary tables S7 to S9) with Meta-PolarIm (Fig. 4c) to evaluate the polarization detection accuracy using a customized measurement setup (Supplementary Fig.S16)[22]. The measurement error $\Delta S_i^j$ (i=1, 2, 3 for the Stokes parameters; j=1,2... 8 for the polarization states) was calculated by subtracting the measurement data from the reference values obtained from theoretical calculation (methods section). The mean absolute error (MAE) for $S_1$, $S_2$, $S_3$ were found less than 5% for all three wavelength bands (Supplementary Table S10 and Figs. S14-16). We also performed statistical analysis for the errors of all pixels in the imaging sensor, including measurement errors for angle of polarization (AOP=$\frac{1}{2}\arctan\frac{S_2}{S_1}$), degree of circular polarization (DOCP=$S_3/S_0$) and degree of linear polarization (DOLP=$\sqrt{S_1^2 + S_2^2}/S_0$ ) for the eight polarization states over the three wavelength bands (Fig. 4d). The results suggested that 90% of the polarimetric imaging pixels has reasonably small measurement errors for DOLP (<3%), AOP (<1.8°) and DOCP (<2% for green and red, <6% for blue) (Supplementary Table S10 and Figs. S17-S19). Our results confirmed that the NIL-based nanomanufacturing method is suitable for producing multi-layer metasurface devices with reasonably high performance and uniformity across centimeter scale.

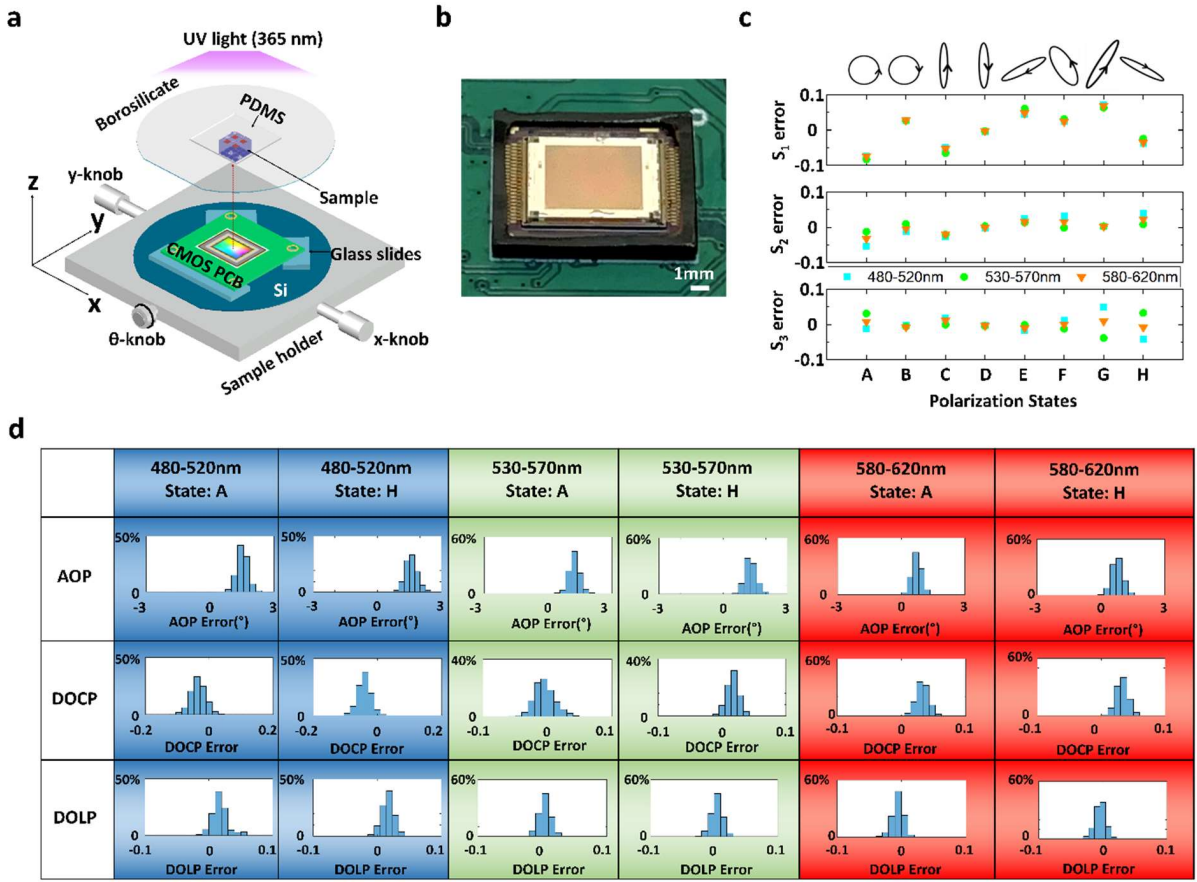

**Fig. 4. Multi-color full-Stokes polarization state detection using metasurface polarimetric imaging sensor. a**, Schematic of integrating metasurface polarization filter arrays onto CMOS imaging sensor. CMOS circuit board is firstly mounted onto 3D rotation stage and leveled, then the metasurface array is aligned and bonded onto the board via UV mask aligner. **b**, An optical image of the integrated metasurface polarimetric imaging sensor. **c**, Error analysis of multi-color full-Stokes parameter detection for eight polarization states (A to H). **d**, Multi-color AOP, DOCP and DOLP detection error distributions of all metasurface pixels for polarization states A and H. X-axes represent the errors and Y axes represent the corresponding percentage of pixels.

**Polarization imaging application.**

The chip-integrated full stokes polarimetric imaging sensors have a broad range of applications. As proof-of-concept demonstrations, here we show the imaging results of several objects, including a plastic fork, a pair of 3D glasses, and a beetle (Fig. 5, measurement setup illustrated in Supplementary Figs. S20). The polarimetric images of the plastic fork (Fig. 5a) and 3D glasses (Fig. 5b) were obtained in transmission mode with ~90° LP as the input light. These objects exhibited poor contrasts from the background in signal intensity (S0) in all color bands; however, the AOP and DOCP images showed distinct contrasts. This was attributed to spatially varying optical birefringence (from local stress) in the plastic fork and the designed polarization response from the glasses. In addition, their DOCP images in the blue and red channels produced visually different polarization signals, indicating the wavelength-dependent polarization response. We also took images of a green June beetle sealed in resin in reflection mode (Fig. 5c). The beetle elytra regions also presented a signature of wavelength dependence in DOCP images, showing right-handed CP signal (DOCP>0) in green channel, left-handed (DOCP<0) in blue channel, but only low-contrast signals in red channel, respectively. The above imaging results demonstrated the unique advantage of our metasurface-integrated Meta-Polarim to enhance imaging contrast by incorporating full-Stokes polarimetric signals in multi-wavelength channels, which are otherwise not available by conventional imaging sensors.

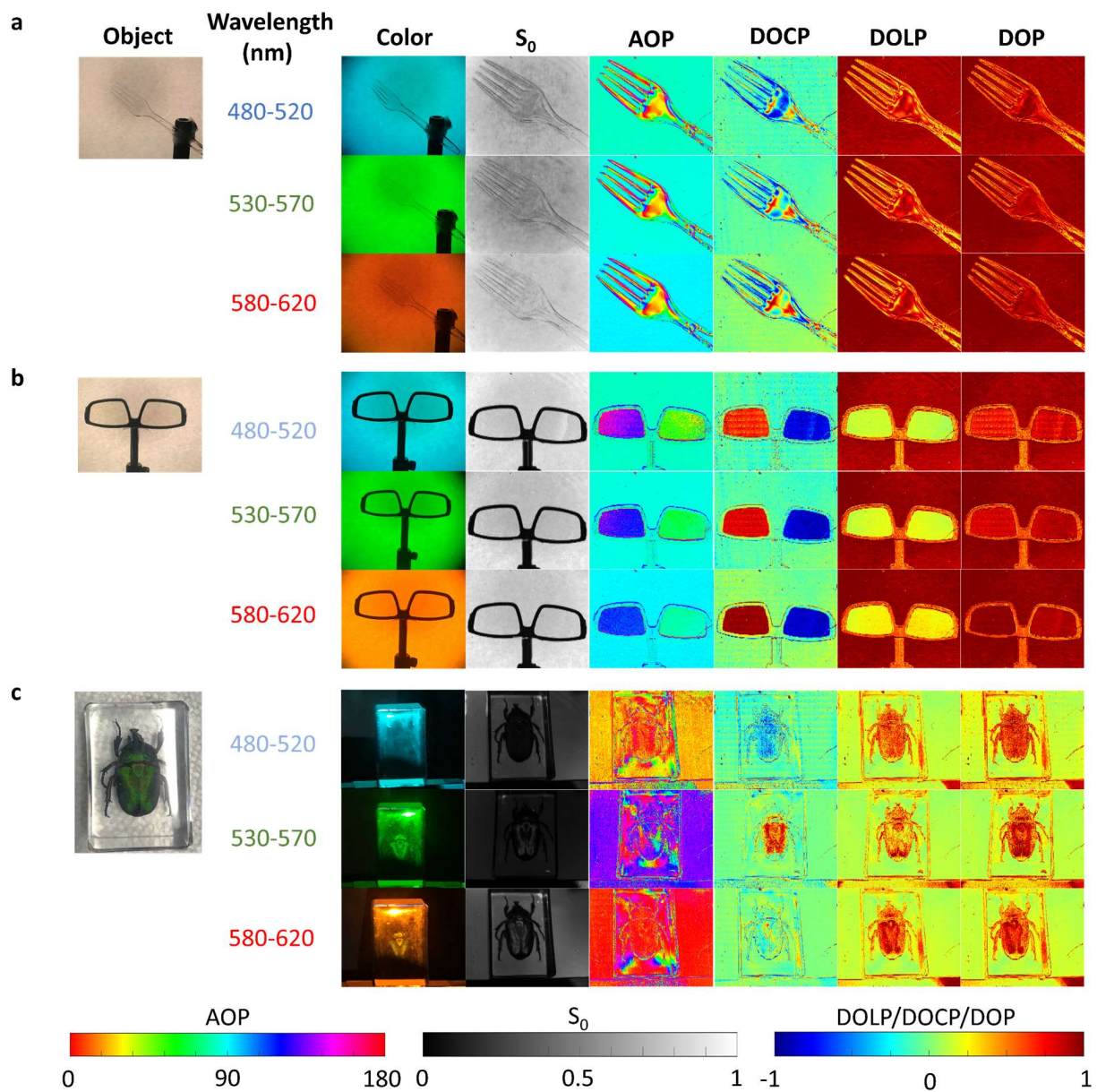

**Fig. 5. Multi-Color full-Stokes polarimetric images of exemplary objects. a, b**, Transmission, color filtered, $S_0$, AOP, DOLP, DOCP, and DOP images of (a) a plastic fork, and (b) a pair of 3D glasses. The images were taken with the LP input as background. **c**, Reflection, color filtered, $S_0$, AOP, DOLP, DOCP, and DOP images of a Green June beetle. All the images were taken with LP input light source and unpolarized white paper as background. Color channels were obtained by applying bandpass filters, i.e. 480 -520 nm for blue, 530-570nm for green, and 580-620nm for red.

## (5) Conclusions

In summary, we propose and demonstrate a scalable manufacturing strategy of multilayer metasurfaces by synergistically co-designing multi-level NIL processes and metasurface structures. In particular, we leveraged the nanometer-scale resolution and 3D pattern transfer capability of NIL to design and fabricate multi-layer dielectric and metallic hybrid nanostructures as metasurface polarization filter arrays (MPFAs). We demonstrate successful fabrication of densely arranged nanostructures (period ~200 nm and critical dimension <100 nm) over ~0.2 cm$^2$ area with uniform and accurate dimension control (linewidth standard deviation less than 4 nm), high interlayer alignment accuracy (translational error ~200 nm, rotation error < 0.017 degrees) and high performance over broad visible wavelength ranges (>200 nm, with a large CPER of >8). We further bonded the MPFAs to a CMOS imager to create a metasurface polarimetric imaging sensor, i.e. Meta-PolarIm, for compact, single-shot, broadband polarimetric imaging in visible wavelengths with high polarization state measurement accuracy (<5%). Most importantly, the presented NIL-based manufacturing process enabled low-cost and high-throughput fabrication of these devices, which is essential for future commercialization and broad deployment in various applications, from biomedical imaging to material defects analysis and remote target detection. This multilayer NIL-metasurface codesign approach can be adapted to the fabrication of many other metasurface structures, enabling a new scalable manufacturing and on-chip integration strategy of metasurface devices and their related optic or optoelectronic systems. By speeding up the prototyping process and enabling low-cost, large-scale production of such devices and systems, our design and manufacturing strategy can inspire future innovations in profound applications from advanced metrology[31], augmented reality[16], and holography to optical computation[32] and energy conservation[33]  that are key to next-generation commercial electronics, national security and sustainability.

## (6) Methods

### *Materials*

Poly(benzyl methacrylate) ($\geq$ 99.0 %), Propylene glycol monomethyl ether acetate ($\geq$ 99.5 %), Pentaerythritol tetra acrylate, Isobutyl methacrylate ($\geq$ 97.0 %), Anisole ($\geq$ 99.7 %), and trichloro($1H,1H,2H,2H$-perfluorooctyl)silane, Octadecyl acrylate ($\geq$ 97.0 %), and $1H,1H,2H,2H$-perfluoro-1-decanol (97 %) were purchased from Sigma-Aldrich. BYK-310 and BYK-3570 were purchased from BYK Additives and Instruments. Omnirad 1173 and Omnirad TPO were purchased from IGM Resins. (Acryloxypropyl) methyl siloxane homopolymer was purchased from Gelest. PMMA (950K A2 and 495K A3) was purchased from MicroChem. AMOPRIME was purchased from AMO GmbH. CN-292, SR-9003-B, and CN-975 were purchased from Satomer. AZ-1505 positive photoresist was purchased from MicroChemicals. Gel-box AD-22AS-00 was purchased from Gel-Pak. All chemicals used as received without further purification.

### *Resist preparation for NIL & CMOS bonding processes*

The thermal NIL resist was prepared by diluting thermoplastic polymer (poly-benzyl methacrylate, or PBMA) in Propylene glycol monomethyl ether acetate (PMA) as solvent, with a small amount of surface additive (BYK-310) added for lowering surface tension. The UV-NIL resist was prepared by mixing (Acryloxypropyl) methyl siloxane homopolymer with cross-linker (Pentaerythritol tetraacylated) photo initiators (Omnirad 1173 and Omnirad TPO) and surface additive (BYK-3570) in Isobutyl methacrylate (IBMA). For the CMOS bonding process, another UV-curable polymer was prepared by mixing fast-reacting, low-viscosity, acrylate oligomers (e.g. SR-9003-B and CN-292), a surface additive ($1H,1H,2H,2H$-Perfluoro-1-decanol BYK-3570), and photo initiators (Omnirad 1173 and Omnirad TPO) into IBMA solvent. All the solutions were stirred overnight at room temperature and filtered before use.

### *Mold fabrication for Si metasurface & VCDGs by EBL*

The Si metasurface mold was fabricated by EBL (Supplementary Fig. S6). A polymethyl methacrylate (PMMA) bi-layer was spin-coated (PS-80, Headway Research Inc.) on a cleaned silicon (Si) substrate (1 mm thick, with 80 nm thermal $SiO_2$) and post-baked 5 min at 180 °C. Then a 10 nm Cr layer was deposited on the PMMA as a discharging layer for EBL by thermal evaporator (Denton Bench Top Turbo, Denton Vacuum, LLC) at a deposition rate of 0.2 Å $s^{-1}$. Then EBL was carried out (ELS-7000, Elionix) with an acceleration voltage of 100 kV, a beam current of 1 nA, a field size of 300 μm with a minimum step size of 5 nm, and an exposure dose of 1200 μC $cm^{-2}$. After EBL, the Cr discharging layer was stripped and the patterns were developed in a 1:3 ratio (v/v) of methyl isobutyl ketone (MIBK)/isopropyl alcohol (IPA) solution for 2 minutes, rinsed in IPA, and dried with nitrogen. Then, a 10 nm Cr layer was deposited by using thermal evaporator followed by an oxygen plasma descum (Tergeo plasma cleaner, 20 W, 10 sccm, 40 s) process. The sample was immersed in remover PG solution for 15 minutes at 80 °C for the lift-off process, rinsed with IPA and then DI water, and dried. The $SiO_2$ layer was etched by reactive-ion-etcher (RIE) (PlasmaTherm 790, $CHF_3$ = 40 sccm, $O_2$ = 3 sccm, 40 mTorr, 250 W) using Cr as a hard mask. Finally, the Cr hard mask was stripped by Cr etchant.

To fabricate NIL mold for the VCDGs, a thick fused silica wafer (6 mm) was chosen as the substrate (Supplementary Fig. S7) to minimize mold bending during NIL. Then fused silica dicing, sample cleaning, EBL writing, development, Cr evaporation, and lift off were carried following the same process mentioned above to produce the nanostructured Cr hard masks. The EBL exposure doses were adjusted for designed structural dimensions. The Cr mask was used to etch 150 nm deep into fused silica by RIE using the same recipe as aforementioned. Differently, a mesa structure (roughly 1.5 $cm^2$, height = 2 μm) was intentionally fabricated in an additional RIE process to better accumulate pressure in the nanopatterned region. The mesa structure provided more uniformly imprinted structures using our imprinter. Both Si and fused silica molds were solvent and RCA-1 cleaned, and they were treated using trichloro(1$H$,1$H$,2$H$,2$H$-perfluorooctyl) silane in the vacuum oven for 30 min at 100 °C to form the self-assembled monolayers (SAMs) on the surface, which acted as an anti-sticking layer during the NIL process[32].

### *Fabrication of Si metasurface in a tri-layer scheme*

First, 130 nm α-Si was deposited on the pre-cleaned fused silica sample using plasma-enhanced CVD (PECVD) (Oxford Plasmalab 100, $SiH_4$ = 480 sccm, 1200 mTorr, 15 W, 350 °C), followed by 60 nm $SiO_2$ deposition using the same tool ($SiH_4$ = 170 sccm, $N_2O$ = 710 sccm, 1000 mTorr, 20 W, 350 °C) without breaking chamber vacuum. After the substrate preparation, a tri-layer structure was employed for the thermal NIL process. Namely, a PMMA layer (950k A2, thickness of 90nm) was spin-coated and post-baked 5 min at 200 °C, followed by evaporation of ~15 nm $SiO_2$ mid-layer (Kurt J. Lesker) at a deposition rate of 0.5 Å $s^{-1}$, and then spin-coating of thermal NIL resist and post-baking (5 min at 180 °C). The thermal NIL was carried out using a nanoimprinter (THU400, Zhenjiang Lehua Electronic Technology Co. Ltd.) at a nominal temperature reading of 55 °C and pressure of 750 KPa for 15 min in vacuum. Then the residual layer was RIE etched by oxygen plasma ($O_2$ = 10 sccm, 10 mTorr, 100 W), where $SiO_2$ mid-layer acted as the etch-stop layer to enable sufficient over etching time for uniform removal of the residual layer. The nanopatterns in the resist were transferred to $SiO_2$ mid-layer by another RIE etching ($CHF_3$ = 25 sccm, $O_2$ = 1 sccm, 10 mTorr, 100 W), and the PMMA bottom layer was RIE etched by oxygen plasma ($O_2$ = 10 sccm, 10 mTorr, 30 W). The high etching selectivity between $SiO_2$ and PMMA is beneficial for reliable patterning in a relatively thick PMMA layer, and helps form a mushroom-like structure in the $SiO_2$/PMMA stack to minimize accumulation of metal on the sidewall of PMMA, which facilitated high-yield lift-off process and minimized feature distortion. The fabricated sample was immersed in remover PG solution for 15 minutes at 80 °C for lift-off, and later rinsed with IPA and DI water, followed by 10 nm Cr layer deposition by thermal evaporation. The 60 nm $SiO_2$ hard mask layer was etched by RIE ($CHF_3$ = 40 sccm, $O_2$ = 3 sccm, 40 mTorr, 250 W) using Cr as a hard mask, and Cr was stripped by chromium etchant. Finally, the 130 nm α-Si film was etched using inductively coupled plasma (ICP) RIE (PlasmaTherm Apex ICP, $Cl_2$ = 100 sccm, Ar = 5 sccm, 10 mTorr, 250 W) using $SiO_2$ as a hard mask to complete Si metasurface fabrication. The $SiO_2$ hard mask layer was left without intentional removal, but its thickness was taken into account of the whole spacer layer thickness calculation.

### *Fabrication of VCDGs using UV-NIL*

The VCDGs were fabricated in a significantly simpler fabrication process. First, adhesion promoter (AMOPRIME) was spin-coated on pre-cleaned fused silica chips (some with Si metasurface for device integration and some others without Si metasurface used as process monitors) and post-baked 10 min at 115 °C on a hot plate. The prepared UV-NIL resist was spin-coated on the substrate, followed by UV-NIL using the fabricated fused silica VCDG mold on mask aligner (MJB4, Suss MicroTec). Three different fringes were visualized on the mask aligner TV monitor for the ease of alignment. Once alignment was verified, 1.5 s UV exposure was used to cross-link the resist, which turned to be a polymer similar to $SiO_x$ in optical index after curing (UV-NIR spectroscopic ellipsometry, J.A. Woollam, M-2000) (Supplementary Fig. S8). The UV resist had a low viscosity to easily fill with relatively low pressure[33]. After UV-NIL, the printed resist scaffold was treated using a mild oxygen plasma process ($O_2$ = 10 sccm, 10 mTorr, 30 W) to activate the hydroxyl groups on the surface [34]. A layer of Cr (2 nm) was evaporated followed by Al deposition at 2.5 Å $s^{-1}$ to form the VCDG gratings. A high vacuum level (1 to $3 \times 10^{-7}$ Torr) was useful to obtaining smoother surface morphology of VCDG (supplementary Fig. S9) by decreasing residual gases in the and reducing contaminants [35]. Finally, a 200 nm $SiO_2$ layer was deposited as an encapsulation layer to avoid further oxidation of the Al surface by using a radio-frequency (RF) sputtering system (Kurt J. Lesker) at a deposition rate of 0.5 Å s$^{-1}$.

### *Vertical alignment and integration of VCDGs on Si metasurface*

Here two sets of gratings with slightly different periods (e.g. $P_1$ of 4 µm on the substrate and from Si metasurface mold, and $P_2$ of 4.2 µm on the VCDG mold) acted as the Moiré marks. The two gratings would produce periodic stripes under illumination, with the period $P_{fringe}$ calculated as $P_{fringe} = P_1 \cdot P_2 / (P_2 - P_1)$ = 84 µm, when the substrate and mold were brought close to each other, e.g. with a small gap less than 10 µm. To minimize the alignment error, four groups of alignment markers (AM$_1$, AM$_2$, AM$_3$, and AM$_4$, respectively) were placed next to the NIL-patterned area, in another word separated by 7.2 mm horizontally and 5.6 mm vertically from each other. Noticeably, our process differ from previously studies that required metal deposition [27, 28], because the large optical index difference from α-Si metasurface ($n$ = 3.58 at 632 nm) to the substrate SiO$_2$ ($n$ = 1.49 at 632 nm) provided distinguishable contrast and eliminated the needs of metallic coating.

### *CMOS bonding process*

The integrated multilayer metasurface chip was diced and bonded onto the customized CMOS sensor as follows. Here AZ-1505 photoresist (PR) was spin-coated on both sides of the fabricated sample and post-baked 1 min at 90 °C as a protection layer during chip dicing, then the sample was diced into 7.2 mm × 5.6 mm rectangular shape using a dicing saw (DAD320, DISCO Corporation). Afterwards, the sample was immersed in acetone to remove PR, rinsed in IPA, and dried with nitrogen blow. A thin PDMS film of ~1mm was detached from a commercially available Gel-box and attached to a 4-inch borosilicate wafer as an intermediate host layer for the diced chip. A customized CMOS sensor was brought together with a printed circuit board (PCB)[31], and mounted onto a customized support by Kapton tape, formed by stacked glass slides taped on a 4-inch Si wafer, to maintain the surface evenness considering that the backside of PCB had protruding electrical components. Then the UV-curable polymer was spin-coated on the CMOS imager, and the CMOS PCB was loaded into the mask aligner (MJB4, Suss MicroTec). After precise alignment the CMOS PCB was moved up in the z-direction and made contact with the metasurface chip, initiating polymer flow. Then the polymer was cross-linked under UV exposure (365 nm, 350 W) for 10 min to ensure appropriate bonding strength.

### *Structural and material characterization*

The linewidth dimension and surface morphology of the α-Si metasurface and the Al VCDGs were inspected by scanning electron microscopy (SEM, Hitachi S-4700 FESEM) with an acceleration voltage of 15 keV and current of 10 µA. A thin layer of Au/Pd was sputtered (Cressington sputter coater 108) on the α-Si metasurface sample to enhance imaging resolution prior to SEM measurements. Optical properties (refractive index $n$, extinction coefficient $k$) of deposited α-Si and $SiO_2$ and cured UV resist films were measured by UV-NIR spectroscopic ellipsometry (J.A. Woollam, M-2000). Olympus BX53 fluorescent microscope coupled Horiba iHR320 imaging spectrometer was utilized to record all the optical images of fabricated samples for the calculation of alignment accuracy. It is noted that the electron microscopy would not be able to effectively detect the α-Si metasurface buried deep under the thick spacer layer (~ 500 nm) after the UV NIL effectively planarized the surface topography. To standardize alignment measurement, the optical images were converted to 8-bit black and white images and processed by setting a color threshold. The transmittance spectra were measured by the same tool, then LPER and CPER were calculated as the previous work [31].

### *Metasurface design and simulation*

The finite-difference time-domain (FDTD) simulations were carried out to calculate transmission efficiency and LPER and CPER of the metal-dielectric hybrid chiral metasurface. All the simulations were conducted with empirically measured optical indexes of each material. Periodic boundary conditions and perfectly matched layers were used within a unit cell along the in-plain direction. The plane wave was applied along the grating width and length direction to calculate LPER and efficiency and super-positioned two orthogonally linearly polarized plane waves were used to represent RCP/LCP light input. The over-hanged structure of the top layer of VCDGs and tilted angle of 6° of the α-Si gratings were considered in the simulations. The mesh sizes were set as 5 nm for higher accuracy[31].

*Reference polarization state value calculation*

Stokes parameters of 16 reference polarization states input were theoretically calculated based on the linear retardance, transmission efficiency, bandwidth of the color filters, the angle of linear polarizer, and super achromatic quarter-wave plate. Firstly, transmission efficiency and linear retardance dispersion data of SAQWP05M-700 (Thorlabs) was obtained from Thorlabs website. Stokes parameter of light transmitted through the linear polarizer with angle of $\theta_1$ and quarter waveplate with fast axis along angle $\theta_2$ can be modelled using the Mueller matrix of a linear diatenuator and a linear retarder:

$$M_{LP} = \frac{1}{2} \times \begin{bmatrix} q+r & (q-r)cos2\theta_1 & (q-r)sin2\theta_1 & 0 \\ (q-r)cos2\theta_1 & (q+r)cos^2 2\theta_1 + \sqrt{qr}sin^2 2\theta_1 & (q+r-2\sqrt{qr})sin2\theta_1 cos2\theta_1 & 0 \\ (q-r)sin2\theta_1 & (q+r-2\sqrt{qr})sin2\theta_1 cos2\theta_1 & (q+r)sin^2 2\theta_1 + \sqrt{qr}cos^2 2\theta_1 & 0 \\ 0 & 0 & 0 & 2\sqrt{qr} \end{bmatrix} \quad (1)$$

Here, $\theta_1$ represents the transmission axis of the linear polarizer, $q\ and\ r$ represents the maximum and minimum transmission efficiency of linear polarizer, as extracted from data provided by Thorlabs website. LPER can be expressed as LPER= $q/r$.

$$M_{retarder} = \begin{bmatrix} 1 & 0 & 0 & 0 \\ 0 & cos^2 2\theta_2 + sin^2 2\theta_2 cos\,\delta & sin2\theta_2 cos2\theta_2(1 - cos\,\delta) & -sin2\theta_2 sin\,\delta \\ 0 & sin2\theta_2 cos2\theta_2(1 - cos\,\delta) & sin^2 2\theta_2 + cos^2 2\theta_2 cos\,\delta & cos2\theta_2 sin\,\delta \\ 0 & sin2\theta_2 sin\,\delta & -cos2\theta_2 sin\,\delta & cos\,\delta \end{bmatrix} \tag{2}$$

Here, $\theta_2$ represents the angle fast axis of the retarder, $\delta$ represent retardance, as extracted from data provided by Thorlabs website.

$$S_{in}^{\lambda} = M_{retarder} \cdot M_{LP} \cdot \begin{bmatrix} 1 \\ 0 \\ 0 \\ 0 \end{bmatrix} \tag{3}$$

It is noteworthy that both $\delta$ and fast axis angle $\theta_2$ are wavelength dependent. Final Stokes parameter was averaged out after we obtained $S_{in}$ at each wavelength accounting for wavelength dependency of $\delta$ and $\theta_2$ using the equation:

$$S_{in} = \frac{\sum_{i=1}^{n} S_{in,i}^{\lambda}}{n} \tag{4}$$

***Device instrument matrix calibration process at red, green, blue colors***

For an arbitrary input polarization state $S_\lambda$ with input wavelength of $\lambda$(nm), the captured intensity of a super-pixel as a vector $I$ can be written as the equation below:

$$I = \begin{bmatrix} s_0^{0G\_out} \\ s_0^{90G\_out} \\ s_0^{135G\_out} \\ s_0^{45G\_out} \\ s_0^{LCP\_RB} \\ s_0^{RCP\_RB} \\ s_0^{LCP\_G} \\ s_0^{RCP\_G} \end{bmatrix} = \begin{bmatrix} m_{11}^{0G} & m_{12}^{0G} & m_{13}^{0G} & m_{14}^{0G} \\ m_{11}^{90G} & m_{12}^{90G} & m_{13}^{90G} & m_{14}^{90G} \\ m_{11}^{135G} & m_{12}^{135G} & m_{13}^{135G} & m_{14}^{135G} \\ m_{11}^{45G} & m_{12}^{45G} & m_{13}^{45G} & m_{14}^{45G} \\ m_{11}^{LCP\_RB} & m_{12}^{LCP\_RB} & m_{13}^{LCP\_RB} & m_{14}^{LCP\_RB} \\ m_{11}^{RCP\_RB} & m_{12}^{RCP\_RB} & m_{13}^{RCP\_RB} & m_{14}^{RCP\_RB} \\ m_{11}^{LCP\_G} & m_{12}^{LCP\_G} & m_{13}^{LCP\_G} & m_{14}^{LCP\_G} \\ m_{11}^{RCP\_G} & m_{12}^{RCP\_G} & m_{13}^{RCP\_G} & m_{14}^{RCP\_G} \end{bmatrix} \times \begin{bmatrix} S_0 \\ S_1 \\ S_2 \\ S_3 \end{bmatrix} = A_\lambda \times S_\lambda \tag{5}$$

Where matrix $A_\lambda$ is wavelength dependent instrument matrix of the metasurface filter array. $S_\lambda$ can be inversely calculated by solving the equation 5:

$$S_\lambda = A_\lambda^{-1} \times I \tag{6}$$

The measurement setup is the same as our previous work [22]. To obtain $A_\lambda$ at red, green, and blue color ranges, we used three color filters (FBH450-40, FBH550-40, FBH600-40) with bandwidths of 40nm to select targeted wavelength range respectively. For each color, 10 pre-known polarization states $S_{\lambda,4\times10}$ was measured by the device respectively to form an intensity matrix $I_{cam,6\times10}$, the instrument matrix $A_\lambda$ can then be obtained using the equation:

$$A_\lambda = I_{cam,6\times10} \times S_{\lambda,4\times10}^{\mathrm{T}}{}^{-1} \tag{7}$$

Here, the rank of $S_{\lambda,4\times10}^{\mathrm{T}}$ should be 4 to make sure $S_{\lambda,4\times10}^{\mathrm{T}}$ is invertible.

***Stokes parameter analysis***

A moving window spatial scanning discussed in our previous work [31] was first applied during the calibration process to increase the imaging resolution of the polarimetric imaging sensor to 671 by 509 pixels. Each pixel is one polarization filter. The measurement result $S_i^j$ is then averaged out from all of pixels: $S_i^j = \frac{\sum_{m=1,n=1}^{o,p} S_i{}^j{}_{m,n}/S_0{}^j{}_{m,n}}{o\times p}$, i=1, 2, 3, j= 1,2... 8, n=671, p=509, where $S_i{}^j{}_{m,n}/S_0{}^j{}_{m,n}$ denotes the normalized Stokes parameters measured by one pixel. Fig 4c shows the measurement error $\Delta S_i^j$ at normal incidence under multi color input, where $\Delta S_i^j$ is defined as: $\Delta S_i^j = S_i^j - S_R{}_i^j$ (i=1, 2, 3, j=1,2... 8),and $S_R{}_i^j$ denotes reference values of input Stokes parameters.

## (7) Acknowledgements

This project was supported in part by the Department of Energy under grant no. DE-EE0008999 and NSF under grant no. 1809997, 1838443, 1847324, and 1947753. The authors acknowledge Roland Himmelhuber and Micro/Nano fabrication center at University of Arizona for sharing their EBL tool for nanoimprint mold fabrication process. The devices were fabricated and characterized in the NanoFab and Eyring Materials Center (EMC) at Arizona State University. Access to the NanoFab and/or EMC was supported, in part, by NSF grant no. ECCS-1542160.

## (8) Conflict of interests

The authors declare no conflict of interest.

## (9) Contributions

S.C. and J.Z. contributed equally to this work. C.W. and Y.Y. conceived the idea and supervised the study. S.C., J.Z. performed layout design, S.C. performed mold fabrication, metasurface fabrication, structural characterization, and data analysis. S.C., J.Z. performed chip integration process, J.Z. carried out photonic simulation, device characterization and data analysis. N.D. supported device characterization and data analysis. S.C., J.Z., Y.Y., and C.W. wrote the manuscript, and all authors contributed to the manuscript.

# (10) References